\documentclass[aps, pra, twocolumn, nofootinbib,notitlepage]{revtex4-2}
\usepackage[hidelinks]{hyperref}
\usepackage{xspace, amsmath}
\usepackage[braket, qm]{qcircuit}
\usepackage{graphicx, csquotes}

\usepackage[a4paper, total={6.75in, 9in}, margin=2cm]{geometry}

\usepackage{natbib}

\usepackage{subcaption}
\usepackage[font=small,labelfont=bf,
justification=justified,
format=plain]{caption}
\usepackage{enumitem, diagbox, multirow}
\setlist[enumerate]{label*=\arabic*.}

\newcommand{\undr}{\rule{1em}{.75pt}}
\newcommand{\iSWAP}{$iSW\!AP$\xspace}
\newcommand{\CX}{$CX$\xspace}
\renewcommand{\H}{$H$\xspace}
\newcommand{\Hxy}{$H_{XY}$\xspace}
\newcommand{\Hyz}{$H_{YZ}$\xspace}
\newcommand{\Cxyz}{$C_{XYZ}$\xspace}
\newcommand{\Czyx}{$C_{ZYX}$\xspace}
\newcommand{\sqrtXX}{$\sqrt{XX}$\xspace}

\newcommand{\clifford}{$\{H,~S,~CX\}$\xspace}

\renewcommand{\sp}{\hspace{.25em}}
\let\1\undefined
\let\2\undefined
\usepackage{outlines}

\begin{document}
\title{A simple method for compiling quantum stabilizer circuits}
\author{Brendan Reid}
\affiliation{Entropica Labs, Singapore}
\email{brendan@entropicalabs.com}
\begin{abstract}
	Stabilizer circuits play an important role in quantum error correction protocols, and will be vital for ensuring fault tolerance in future quantum hardware. While stabilizer circuits are defined on the Clifford generating set, \clifford,  not all of these gates are native to quantum hardware. As such they must be compiled into the native gateset, with the key difference across hardware archetypes being the native two-qubit gate.
	
	Here we introduce an intuitive and accessible method for Clifford gate compilation. While multiple open source solutions exist for quantum circuit compilation, these operate on arbitrary quantum gates. By restricting ourselves to Clifford gates, the compilation process becomes almost trivial and even large circuits can be compiled manually. 
	
	The core idea is well known: if two Clifford circuits conjugate Paulis identically, they are equivalent. Compilation is then reduced to ensuring that the instantaneous Pauli conjugation is correct for each qubit at every timestep. This is Tableaux Manipulation, so called as we directly interrogate stabilizer tableaux to ensure correct Pauli conjugation. We provide a brief explanation of the process along with a worked example to build intuition; we finally show some comparisons for compiling large circuits to open source software, and highlight that this method ensures a minimal number of quantum gates are employed. 
\end{abstract}
\maketitle\newpage
	\section{Introduction}
	In the future running a quantum algorithm on a fault tolerant machine will not be unlike running a circuit through a quantum API today. Quantum software will ensure that the complexities of fault tolerance --- codes, decoders and logical primitives --- are abstracted away from the end user. To achieve this there are a number of stages that must be automated. 
	
	At the highest level, converting an algorithm of $N$ logical qubits into a network of $M\gg N$ physical qubits will require insight into the algorithm being executed. Logical qubit placement will depend on (logical) two-qubit interactions, and minimising the (logical) ancillae overhead \cite{Litinski2019gameofsurfacecodes}. Once this task is complete the physical circuit must be constructed. Qubit preparation and logical gates, as well as the fundamental operations of the target error correction scheme (e.g. in the case of the surface code, lattice surgery operations such as growth, merge, split etc. \cite{Horsman_2012}) all require a number of rounds of syndrome extraction to ensure errors are caught and accounted for. The result is an extremely deep circuit across an array of physical qubits, with the bulk of operations likely defined in terms of the Clifford generating set \clifford. The final step will be conversion from this gateset into the hardware instruction set. At this final stage the scope for minimising error events is limited, as the noisest processes in the experiment (two-qubit gates and measurements) have already been determined by the previous stages of the pipeline. Nevertheless, we will show that even modest circuit optimisations can have a positive impact on experiment performance.
	
	Quantum hardware has advanced significantly in recent years, and continues to do so. The race to quantum advantage has resulted in a number of public and private enterprises employing varying quantum hardware modalities. The platforms that have achieved the most experimentally are superconducting qubits \cite{GoogleScaling}, trapped ion systems \cite{Quantinuum} and more recently neutral atoms \cite{neutralatoms}. Even within a qubit modality, differences between experimental groups abound. Beyond the disparities of noise models and connectivity, a key difference is in the entangling operations. 
	
	Trapped ion models employ a M\o lmer-S\o rensen interaction, which manifests as a $XX(\theta)=e^{-\frac{i\theta}{2}X\otimes X}$ rotation, where setting $\theta=\pi/2$ makes this interaction maximally entangling \cite{MSgate}. In this work, when speaking of a M\o lmer-S\o rensen gate, we mean $XX(\theta=\pi/2)$ which we denote \sqrtXX. Superconducting systems developed by IBM use the echoed cross-resonance interaction ($ECR$) \cite{ECRGate}. This gate implements $\frac{1}{\sqrt{2}}\left(I\otimes X-X\otimes Y\right)$ onto the qubits, and is equivalent to a \CX gate up to single qubit rotations. Conversely, Google's superconducting qubit processors have either an $iSWAP=e^{\frac{i\pi}{4}\left(X\otimes X+Y\otimes Y \right)}$ or $CZ$ gate as the native two-qubit gate \cite{Arute2019}. 
	
	This brief work will present a method for rewriting quantum stabilizer circuits into alternative sets of Clifford gates, with a focus on circuits with practical applications. We are referring to this process as ``compilation'', although it could equivalently be called gate decomposition. We are not setting ourselves the task of amending a circuit to fit a specific qubit connectivity, i.e. we will not be adding or removing qubits. These restrictions create a narrow problem statement, but one that is key to using quantum computers effectively.

	The process is straightforward and surprisingly powerful when compared to existing available open-source compilers. Importantly this method can be implemented with little effort; it is simple to compile a circuit consisting of 10-20 qubits manually, if a little tedious. There is minimal information to keep track of, as this method only relies upon the instantaneous Pauli conjugation of an individual qubit at each layer of the circuit. We employ an example driven approach to build intuition for the method. We also draw comparison to the Qiskit open-source compilation functionality and show that, for a key class of circuits, we can always reduce the number of quantum operations executed.

	Much work has been done within the Clifford set, from generating random circuits \cite{Bravyi_2021, berg2021simple} to optimizing the number of two-qubit unitaries required within a circuit \cite{Bravyi2021,Maslov2013}. The techniques used in this work are not new; ultimately we are simply tracking and amending the instantaneous Pauli conjugation of individual qubits. The author could not find any work in the literature with significant overlap to the the results contained within, but welcomes indication toward existing work.
	
\section{Preliminaries}

We assume familiarity with quantum computing and quickly summarise the key concepts relevant to this work. The Pauli group consists of  four operators $\{I,~X,~Y,~Z\}$ 
and is defined as the lowest level of the Clifford hierarchy \cite{Gottesman1999}. A Pauli product is a collection of Pauli operators applied to individual qubits: $X_iY_jZ_k$ applies Pauli-$X$ to qubit $i$, $Y$ to qubit $j$ and $Z$ to qubit $k$. Any qubit index not involved in the Pauli product has the identity applied.

The next level of the hierarchy, the Clifford group, constitutes a key component of fault tolerant universal quantum computation \cite{gottesman1997stabilizer}. The Clifford group contains operators that conjugate Paulis into Paulis, and its generating set of operators consists of \clifford. For example, the conjugation of Paulis by the Hadamard gate $H$ is:
\[H^\dagger X H = Z,~~H^\dagger Y H = -Y,~~H^\dagger Z H = X.~\]

A stabilizer circuit consists solely of elements from the Clifford group and destructive operations in the Pauli-$Z$ basis. A \emph{unitary} stabilizer circuit does not contain measurements or reset operations. This unitary circuit can be  defined by how it conjugates Pauli products, and this information can be summarised via \emph{tableaux}.

%

%
	 Consider the following tableau, representing a $CX$ gate where qubit 0 is the control and qubit 1 is the target:
	\begin{equation}\label{tableau:cx}		
		\vcenter{
		\Qcircuit @C=.5em @R=.5em @!R {
			\lstick{0} &\ctrl{1}&\qw\\
			\lstick{1} & \targ &\qw
			}
		}
~=~
	\begin{tabular}{c|cc}
		& $X_0~Z_0$ & $X_1~Z_1$ \\
		\hline
		$\pm$ & $+~+$ & $+~+$ \\
		0 & $X~Z$ & $ \undr~Z$ \\
		1 & $X~\undr$ & $X~Z$
	\end{tabular}.
	\end{equation}
	The columns $X_j$, $Z_j$ detail the conjugation of these Paulis by the circuit. For example, $X_0$ is conjugated by a $CX$ gate into $+X_0X_1$; $Z_1$ is conjugated into $+Z_0Z_1$.  An underscore indicates that the input Pauli does not propagate onto that output wire: a $Z$ ($X$) term commutes through a $CX$ gate if it is input on the control (target) qubit.	Tableaux such as this (or in the alternative Aaronson-Gottesman format) completely define a Clifford interaction, as input Pauli products can be conjugated by first decomposing to their generators and using the distributive property:
	\begin{align}
	\nonumber CX^{\dagger}\!\left(Y_0Z_1\right)CX=&~ CX^{\dagger}\!\left(X_0Z_0Z_1\right)CX\\
	\nonumber=&~\left(X_0X_1\right)\left(Z_0\right)\left(Z_0Z_1\right)\\
	=&~X_0Y_1.
	\end{align}	Where we ignore global phase.
	Importantly unitary stabilizer circuits are also \emph{uniquely} defined via their tableaux. In Ref.~\cite{Aaronson_2004}, Lemma 1 states:
	\begin{displayquote}
		Let $\mathcal{C}_1$, $\mathcal{C}_2$ be unitary stabilizer circuits, and let $\mathcal{T}_1$, $\mathcal{T}_2$ be their respective final tableaux when we run them on the standard initial tableau. Then $\mathcal{C}_1\equiv\mathcal{C}_2$ iff $\mathcal{T}_1\equiv \mathcal{T}_2$. 
	\end{displayquote}
	Where here the ``standard initial tableau" defines all qubits prepared in the Pauli-$Z$ basis, which we assume implicitly throughout. This Lemma tells us that as long as our tableaux are equivalent, the unitary action will be the same. By observing the differences in two tableaux, we are able to directly modify the circuit in a minimal way to achieve the correct Pauli product conjugation.

	In addition to the conjugation of Pauli terms, we will also speak of the \emph{propagation} of Paulis. Propagation simply refers to the qubit indices present in the Pauli conjugation, i.e. $X_0Z_1Y_2$ has the same propagation as $Y_0X_1Z_2$. This is a useful abstraction when thinking about the action of Paulis in circuits, specifically their commutation relations with two-qubit gates. If qubits $j$ and $k$ only interact once and $k$ does not appear in the conjugation of $P_j$, then the instantaneous conjugation\footnote{`Instantaneous conjugation' here referring to the conjugation of a Pauli at a certain point in the circuit.} of $P_j$ must commute through the two-qubit gate acting on $j$ and $k$.
	
	We will employ a number of single qubit Clifford gates to aid intuition when converting between Pauli bases. The nomenclature for these gates is borrowed directly from the \texttt{stim} package \cite{gidney2021stim}. 
 We list them here along with their decomposition into $S$, $\sqrt{X}$ and Pauli gates:
	\begin{outline}
		\1 \H: standard Hadamard gate, swaps $X$ and $Z$ bases.
			\2  $\vcenter{\Qcircuit @C=.5em @R=.2em @!R{\lstick{} & \gate{H} & \qw}} \equiv \vcenter{\Qcircuit @C=.5em @R=.2em @!R{\lstick{} & \gate{S} & \gate{\sqrt{X}} & \gate{S}}}$
		\1 \Hxy: a Hadamard-like gate that swaps $X$ and $Y$ bases.
			\2 $\vcenter{\Qcircuit @C=.5em @R=.2em @!R{\lstick{} & \gate{H_{XY}} & \qw}} \equiv \vcenter{\Qcircuit @C=.5em @R=.2em @!R{\lstick{} & \gate{S} & \gate{Y}}}$
		\1 \Hyz: a Hadamard-like gate that swaps $Y$ and $Z$ bases.
			\2 $\vcenter{\Qcircuit @C=.5em @R=.2em @!R{\lstick{} & \gate{H_{YZ}} & \qw}} \equiv \vcenter{\Qcircuit @C=.5em @R=.2em @!R{\lstick{} & \gate{\sqrt{X}} & \gate{Z}}}$
		\1 \Cxyz: a basis-cycle gate, moves $X\rightarrow Y \rightarrow Z$.
			\2 $\vcenter{\Qcircuit @C=.5em @R=.2em @!R{\lstick{} & \gate{C_{XYZ}} & \qw}} \equiv \vcenter{\Qcircuit @C=.5em @R=.2em @!R{\lstick{} & \gate{\sqrt{X}} & \gate{S}}}$
		\1 \Czyx: a reverse basis-cycle gate, moves $Z\rightarrow Y \rightarrow X$.
			\2 $\vcenter{\Qcircuit @C=.5em @R=.2em @!R{\lstick{} & \gate{C_{ZYX}} & \qw}} \equiv \vcenter{\Qcircuit @C=.5em @R=.2em @!R{\lstick{} & \gate{S} & \gate{\sqrt{X}} & \gate{Z}}}$
	\end{outline}
	The role of Pauli gates in these circuits is to ensure that the sign changes are consistent on both sides of each equivalency. For example, \Hxy conjugates Pauli-$Z$ like $H_{XY}^\dagger Z H_{XY}=-Z$, whereas an $S$ gate conjugates: $S^\dagger Z S = +Z$. As such, \Hxy and $S$ are equivalent up to the application of a Pauli gate. In practice this equivalency can be tracked in software and we refer to it as the Pauli frame.

The \CX and \iSWAP gates are typically used to classify the different kinds of entangling Cliffords. Gates such as $CZ$, \sqrtXX and $ECR$ are classed as \CX-like, as they are within single qubit rotations to \CX. Gates such as $CXSW\!AP$ and $CZSW\!AP$ (again borrowing terminology from the \texttt{stim} package) are \iSWAP-like, as they are within single qubit rotations to \iSWAP. 
 Throughout we will focus on compiling \CX-like gates into \CX-like gates, although the process outlined below is equally valid for compiling \iSWAP-like into \iSWAP-like. In Appendix \ref{iswap_discussion} we provide discussion around exchanging entangling Clifford type: compiling a \CX-like gate into an \iSWAP-like gate and vice versa. We omit entirely discussion around the two-qubit, non-entangling Clifford gate $SW\!AP$. 
 

\section{Tableaux Manipulation}\label{sec:tm}
	A common approach to circuit compilation is to identify repeated components of the circuit and compile these individually, before replacing them back into the circuit in the relevant locations. From here, gate optimisations and cancellations can take place to reduce the number of operations and the circuit depth. When our circuits are composed solely of Clifford operations, we can instead tackle the compilation of the entire circuit at once. 
	
	Consider a circuit $\mathcal{C}$ that we wish to compile into a gateset $G$. This circuit will provide us with a desired (or `target') Pauli conjugation on each qubit: $\bar{P}_j =\mathcal{C}^{-1}P_j\mathcal{C}$ where $P\in\{X,Z\}$. Rather than working on $\mathcal{C}$ directly, we can instead create a new circuit $\mathcal{C}^\prime$ whose only entries are gates in $G$. This initial condition provides us with conjugations $P_j^\prime=(\mathcal{C}^\prime)^{-1}P_j\mathcal{C}^\prime$. From here the task becomes: what gates do I need to add to $\mathcal{C}^\prime$ to make $P^\prime_j\equiv \bar{P}_j$ for all $j$?
	
	Motivated by the disparities of native entangling Cliffords in different quantum hardware, we employ the following initial condition for our compiled circuit $\mathcal{C}^\prime$: for each entangling operation in $\mathcal{C}$, introduce an entangling operation native to $G$ into $\mathcal{C}^\prime$. A circuit consisting solely of entangling operations is the shortest possible circuit after compilation. From this starting point, we can work through qubits individually and ensure that their Pauli conjugation is correct. 
	
	Tableaux Manipulation has two key steps: first fix the \emph{propagation} of Paulis in the circuit, and then fix the \emph{conjugation} of Paulis after the circuit. To fix the propagation, the process is as follows: 
	\begin{enumerate}[label*=\arabic*.]
		\item For each qubit index $j$, isolate its subcircuit (the timeline of gates it is involved in)
		\item For each entangling operation between qubits $j$ and $k$, determine if $k$ is in the propagation of the target $\bar{P}_j$:
		\begin{enumerate}[label*=\arabic*.]
		\item If it is, add a gate to the circuit (prior to the entangling operation) such that the instantaneous Pauli conjugation of $P_j$ \emph{anticommutes} with the two-qubit gate.
		\item If it is not, add a gate to the circuit (prior to the entangling operation) such that the instantaneous Pauli conjugation of $P_j$ \emph{commutes} with the two-qubit gate.
		\end{enumerate}	
	\end{enumerate} 
	With the caveat that, if the instantaneous conjugation already anticommutes with the two-qubit interaction (or commutes, as the case may be) then no gate need be added. 
	
	Once the propagation of all $P^\prime$ matches that of $\bar{P}$, we must fix the conjugation. For each qubit $j$, we compare the output on qubit $j$ for both $\bar{P}_j$ and  $P^\prime_j$\footnote{This is equivalent to inspecting the diagonal entries of the tableaux we are comparing.}. Correcting the conjugation is then choosing the appropriate basis-change Pauli from a lookup table, Table~\ref{table:conjugation_fixes}.
	\begin{table*}[t]
		\setlength{\tabcolsep}{7.5pt} 
		
	\begin{tabular}{cc|llccrr}
 \hline\\
		\multicolumn{2}{c|}{} & \multicolumn{2}{c}{$\bar{X}=X$} & \multicolumn{2}{c}{$\bar{X}=Y$} & \multicolumn{2}{c}{$\bar{X}=Z$}  \\  
		\multicolumn{2}{c|}{\multirow{-2}{*}{\diagbox{$P^\prime$}{\hspace{2em}$\bar{P}$}}} & $\bar{Z}=Y$& $\bar{Z}=Z$& $\bar{Z}=X$& $\bar{Z}=Z$ & $\bar{Z}=X$& $\bar{Z}=Y$\\ \hline
  \\
		\multirow{2}{*}{$X^\prime=X$}& $Z^\prime=Y$  & $I$  & \Hyz & \Hxy & \Cxyz & \Czyx & $H$  \\ 
		& $Z^\prime=Z$ & \Hyz  & $I$  & \Cxyz & \Hxy & $H$ & \Czyx  \\ \\
		\multirow{2}{*}{$X^\prime=Y$}& $Z^\prime=X$& \Hxy & \Czyx    & $I$ &  $H$ & \Hyz & \Cxyz\\ 
		& $Z^\prime=Z$ & \Czyx &  \Hxy  & $H$& $I$ &\Cxyz & \Hyz          \\
		\\
		\multirow{2}{*}{$X^\prime=Z$} &$Z^\prime=X$& \Cxyz & $H$  &  \Hyz  &          \Czyx  & $I$ & \Hxy\\ 
		& $Z^\prime=Y$ &  $H$  &  \Cxyz  & \Czyx   & \Hyz & \Hxy & $I$ \\
  \hline
	\end{tabular}
	\caption{Lookup table for changing Pauli basis depending on the desired conjugation $\bar{P}$ and the current conjugation $P^\prime$. For example if our desired conjugation is $(\bar{X},~\bar{Z}) = (Y,~X)$ and our current conjugation is $(X^\prime,~Z^\prime) = (Z,~Y)$, we apply a \Czyx gate to that qubit.}
	\label{table:conjugation_fixes}
	\end{table*}

\subsection{Worked Example}
Here we work through an example of using the Tableau Manipulation method to compile a 4-qubit system, ostensibly measuring a $XX$ and $ZZ$ stabiliser with two auxiliary qubits:
\begin{align}\label{tableau:interlaced}
	&\nonumber\vcenter{
		\Qcircuit @C=.5em @R=.5em @!R {
			\lstick{0} & \gate{H} & \ctrl{1} & \ctrl{2} & \qw  & \gate{H} & \qw \\
			\lstick{1} & \qw & \targ & \qw & \ctrl{2} &\qw &\qw  \\
			\lstick{2} & \qw & \ctrl{1} & \targ & \qw &\qw &\qw \\
			\lstick{3} & \qw & \targ & \qw & \targ &\qw &\qw \\
		}
	}
	\\
	\\
	&\sp=\nonumber\begin{tabular}{c|cccc}
		& $X_0~Z_0$ & $X_1~Z_1$ & $X_2~Z_2$ & $X_3~Z_3$\\
		\hline
		$\pm$ & $+~+$ & $+~+$ & $+~+$& $+~+$\\
		0 & $X~Z$ & $ \undr~X$ & $ \undr~X$ & $ \undr~X$ \\
		1 & $\undr~X$  & $X~Z$ & $\undr~\undr$  & $\undr~Z$ \\
		2 & $\undr~X$ & $\undr~\undr$  & $X~Z$ & $\undr~Z$ \\
		3 & $\undr~X$ & $X~\undr$  & $X~\undr$  &  $X~Z$ \\
	\end{tabular}.
\end{align}
For this exercise we'll use an $ECR$ gate, common to some superconducting systems. We represent it with the following circuit diagram and display it's tableau:
\begin{eqnarray}\label{tableau:ecr}	
	\vcenter{
		\Qcircuit @C=.5em @R=.5em @!R {
			\lstick{0} & \gate{{ECR}_{0}}\qwx[1] & \qw \\ 
			\lstick{1} & \gate{ECR_1} & \qw }}
	& \sp = \sp &
	\begin{tabular}{c|cc}
		& $X_0~Z_0$ & $X_1~Z_1$ \\
		\hline
		$\pm$ & $-~-$ & $+~+$ \\
		0 & $Y~Z$ & $ \undr~Z$ \\
		1 & $X~\undr$ & $X~Y$
	\end{tabular}.
\end{eqnarray}As the $ECR$ gate is not symmetric we must keep track of which qubit is the control and target during each interaction. Similar to a \CX gate, Pauli-$Z$ commutes through on the `control' wire (0) and Pauli-$X$ commutes through on the `target' wire (1).

We employ our initial condition from Tableaux Manipulation, and create a new circuit whose only entries are ECR:
\begin{align}\label{tableau:interlaced_ecr}
	&\nonumber\vcenter{
		\Qcircuit @C=.5em @R=.5em @!R {
			\lstick{0} & \qw & \gate{ECR_0}\qwx[1] & \gate{ECR_0}\qwx[2] & \qw  & \qw  &\qw\\
			\lstick{1} & \qw & \gate{ECR_1} & \qw & \qw& \gate{ECR_0}\qwx[2] &\qw   \\
			\lstick{2} & \qw & \gate{ECR_0}\qwx[1] & \gate{ECR_1} & \qw &\qw &\qw \\
			\lstick{3} & \qw & \gate{ECR_1} & \qw & \qw&\gate{ECR_1} &\qw \\
		}
	}
	\\
	\\
	&\sp=\nonumber\begin{tabular}{c|cccc}
		& $X_0~Z_0$ & $X_1~Z_1$ & $X_2~Z_2$ & $X_3~Z_3$\\
		\hline
		$\pm$ & $-~+$ & $-~+$ & $+~-$& $+~-$\\
		0 & $X~Z$ & $ \undr~Z$ & $ Z~Z$ & $ \undr~Z$ \\
		1 & $Y~\undr$  & $Y~X$ & $\undr~\undr$  & $\undr~Z$ \\
		2 & $X~\undr$ & $\undr~\undr$  & $Z~Y$ & $\undr~Y$ \\
		3 & $X~\undr$ & $X~X$  & $X~\undr$  &  $X~Z$ \\
	\end{tabular}.
\end{align}
First we must fix the propagation for each $X_j$, $Z_j$, starting with $j=0$. Note that our desired conjugation $\bar{Z}_0 = Z_0X_1X_2X_3$ has the same propagation as our current $X^\prime_0 = X_0Y_1X_2X_3$, and vice versa. In this case we can simply prepend a Hadamard gate to qubit 0 to swap the action of $X$ and $Z$ Paulis. 

Skipping ahead to qubit 1, we see that the propagation of $\bar{X}_1$ and $X^\prime_1$ are equal, and the only erroneous term is $X_3$ in the conjugation $Z^\prime_1$. Qubit 1 is initially acting as a target qubit in an $ECR$ gate with qubit 0, then as a control qubit in an $ECR$ gate with qubit 3. We need the instantaneous Pauli conjugation of $Z_1$ to commute through this last $ECR$ gate. Below, we track the conjugation where the Paulis on the right hand side represent the conjugation after the operation on the left.
\begin{align*}
	\mathrm{Initial:}&~Z_1\\
	ECR(0,1):&~Z_0Y_1\\
	ECR(1, 3):&~Z_0X_1X_3\\
\end{align*}
As the instantaneous conjugation of $Z_1$ prior to the $ECR(1,3)$ gate is a Pauli-$Y$, we have to convert this to a Pauli-$Z$ in order to commute through. Therefore, we add a \Hyz after $ECR(0,1)$. The circuit now looks like:
\begin{align}\label{tableau:interlaced_ecr_2}
	&\nonumber\vcenter{
		\Qcircuit @C=.5em @R=.5em @!R {
			\lstick{0} & \qw & \gate{H} & \gate{ECR_0}\qwx[1] & \qw & \gate{ECR_0}\qwx[2] & \qw  & \qw  &\qw\\
			\lstick{1} & \qw & \qw & \gate{ECR_1} & \gate{H_{YZ}} & \qw& \gate{ECR_0}\qwx[2] &\qw   \\
			\lstick{2} & \qw & \qw & \gate{ECR_0}\qwx[1] & \qw  &\gate{ECR_1} & \qw &\qw &\qw \\
			\lstick{3} & \qw & \qw & \gate{ECR_1} & \qw & \qw&\gate{ECR_1} &\qw \\
		}
	}
	\\
	\\
	&\sp=\nonumber\begin{tabular}{c|cccc}
		& $X_0~Z_0$ & $X_1~Z_1$ & $X_2~Z_2$ & $X_3~Z_3$\\
		\hline
		$\pm$ & $+~+$ & $+~+$ & $+~-$& $+~-$\\
		0 & $Z~X$ & $ \undr~Z$ & $ Z~Z$ & $ \undr~Z$ \\
		1 & $\undr~Y$  & $Y~Z$ & $\undr~\undr$  & $\undr~Z$ \\
		2 & $\undr~X$ & $\undr~\undr$  & $Z~Y$ & $\undr~Y$ \\
		3 & $\undr~X$ & $X~\undr$  & $X~\undr$  &  $X~Z$ \\
	\end{tabular}.
\end{align}
For qubit 2, the propagations of $\bar{Z}_2$ and $Z^\prime_2$ are equal, and the only erroneous term is the $Z_0$ term in $X^\prime_2$. Again, let's look at the instantaneous Pauli conjugation of qubit 2:
\begin{align*}
	\mathrm{Initial:}&~X_2\\
	ECR(2,3):&~Y_2X_3\\
	ECR(0,2):&~Z_0Z_2X_3\\
\end{align*}
To avoid picking up the $Z_0$ term in the $ECR(0,2)$ gate, we need $Y_2$ to commute through. As qubit 2 is a target in this gate we must convert $Y_2$ into $X_2$:
\begin{align}\label{tableau:interlaced_ecr_3}
	&\nonumber\vcenter{
		\Qcircuit @C=.5em @R=.5em @!R {
			\lstick{0} & \qw & \gate{H} & \gate{ECR_0}\qwx[1] & \qw & \gate{ECR_0}\qwx[2] & \qw  & \qw  &\qw\\
			\lstick{1} & \qw & \qw & \gate{ECR_1} & \gate{H_{YZ}} & \qw& \gate{ECR_0}\qwx[2] &\qw   \\
			\lstick{2} & \qw & \qw & \gate{ECR_0}\qwx[1] & \gate{H_{XY}}  &\gate{ECR_1} & \qw &\qw &\qw \\
			\lstick{3} & \qw & \qw & \gate{ECR_1} & \qw & \qw&\gate{ECR_1} &\qw \\
		}
	}
	\\
	\\
	&\sp=\nonumber\begin{tabular}{c|cccc}
		& $X_0~Z_0$ & $X_1~Z_1$ & $X_2~Z_2$ & $X_3~Z_3$\\
		\hline
		$\pm$ & $+~+$ & $+~+$ & $-~+$& $+~+$\\
		0 & $Z~X$ & $ \undr~Z$ & $ \undr~Z$ & $ \undr~Z$ \\
		1 & $\undr~Y$  & $Y~Z$ & $\undr~\undr$  & $\undr~Z$ \\
		2 & $\undr~X$ & $\undr~\undr$  & $X~Y$ & $\undr~Y$ \\
		3 & $\undr~X$ & $X~\undr$  & $X~\undr$  &  $X~Z$ \\
	\end{tabular}.
\end{align}
Now that the propagation is correct across all inputs, we have to modify the conjugation after the circuit. Where fixing the propagation involved inspecting the columns of the tableau, fixing the conjugations requires inspecting the rows of the tableau. For each $(X^\prime_j,~Z^\prime_j)$ we isolate the output on qubit $j$. For example, $(X^\prime_0,~Z^\prime_0) = (Z_0,~X_0)$ in Eq.~(\ref{tableau:interlaced_ecr_3}). Looking at Eq.~(\ref{tableau:interlaced}) we see this should be $(X_0,~Z_0)$ and so we append a Hadamard onto qubit 0. Repeating this process for qubits 1, 2 and 3 we see that we need to apply a \Hxy to qubit 1, \Hyz to qubit 2 and qubit 3 remains untouched. Skipping ahead to the final circuit in terms of $S$ and $\sqrt{X}$: 
\begin{widetext}
	\begin{equation}
		\vcenter{
			\Qcircuit @C=.5em @R=.5em @!R {
				\lstick{0} & \qw & \gate{S}& \gate{\sqrt{X}}& \gate{S} & \gate{ECR_0}\qwx[1] & \qw & \gate{ECR_0}\qwx[2] & \qw&\gate{S}& \gate{\sqrt{X}}& \gate{S}\\
				\lstick{1} & \qw & \qw& \qw& \qw & \gate{ECR_1} & \gate{\sqrt{X}} & \qw& \gate{ECR_0}\qwx[2]  &\gate{S}&\qw&\qw  \\
				\lstick{2} & \qw &\qw &\qw & \qw & \gate{ECR_0}\qwx[1] & \gate{S} &\gate{ECR_1} & \qw &\qw &\gate{\sqrt{X}} & \qw \\
				\lstick{3} & \qw & \qw&\qw& \qw  & \gate{ECR_1}  & \qw& \qw\qw&\gate{ECR_1} &\qw &\qw&\qw\\
			}
		}\sp=\sp\begin{tabular}{c|cccc}
		& $X_0~Z_0$ & $X_1~Z_1$ & $X_2~Z_2$ & $X_3~Z_3$\\
		\hline
		$\pm$ & $+~+$ & $+~+$ & $+~-$& $+~-$\\
		0 & $X~Z$ & $ \undr~X$ & $ \undr~X$ & $ \undr~X$ \\
		1 & $\undr~X$  & $X~Z$ & $\undr~\undr$  & $\undr~Z$ \\
		2 & $\undr~X$ & $\undr~\undr$  & $X~Z$ & $\undr~Z$ \\
		3 & $\undr~X$ & $X~\undr$  & $X~\undr$  &  $X~Z$ \\
		\end{tabular}.
	\end{equation}
\end{widetext}
This circuit requires a frame correction of $X_2$ to be exactly equivalent to Eq.~(\ref{tableau:interlaced}).
\begin{figure}[t]
	\centering
	\includegraphics[width=\columnwidth]{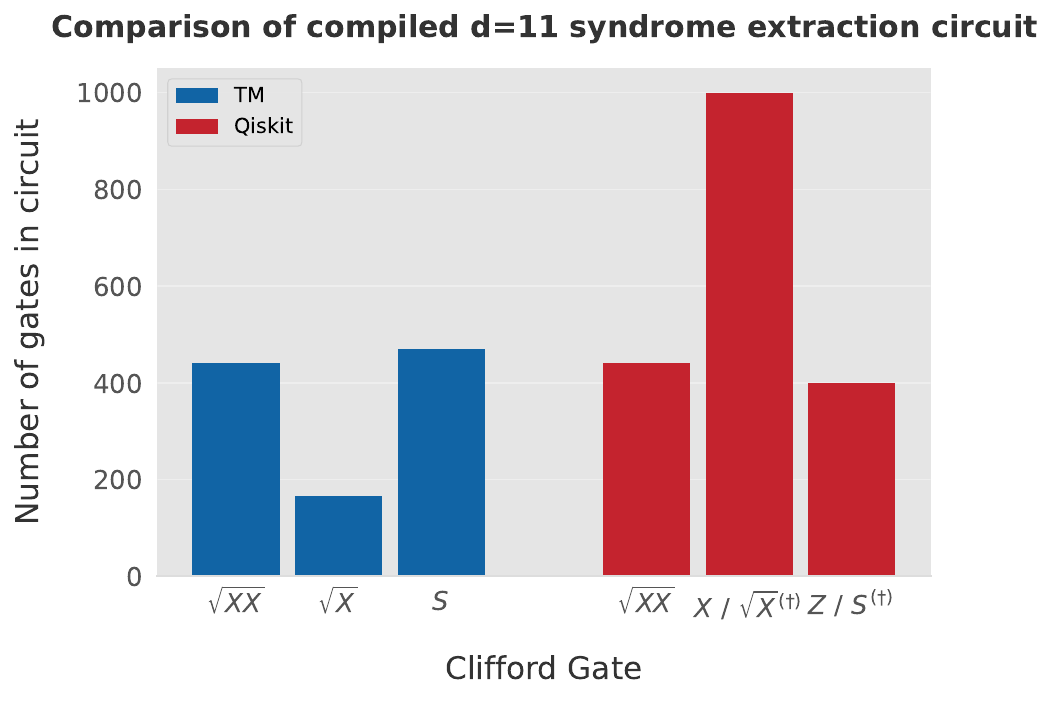}
	\caption{Gate distribution when compiling a syndrome extraction circuit on a rotated surface code of distance $d=11$. The original circuit is 6 layers of gates, defined in terms of $H$ and \CX. Tableaux Manipulation (left, dark blue) and Qiskit (right, dark red) both compiled into a gateset with \sqrtXX as the native two-qubit gate.}
	\label{fig:histogram_sqrtxx}
\end{figure}
\begin{figure*}[t]
	\centering
	\begin{subfigure}{.5\textwidth}
		\centering
		\includegraphics[scale=.5]{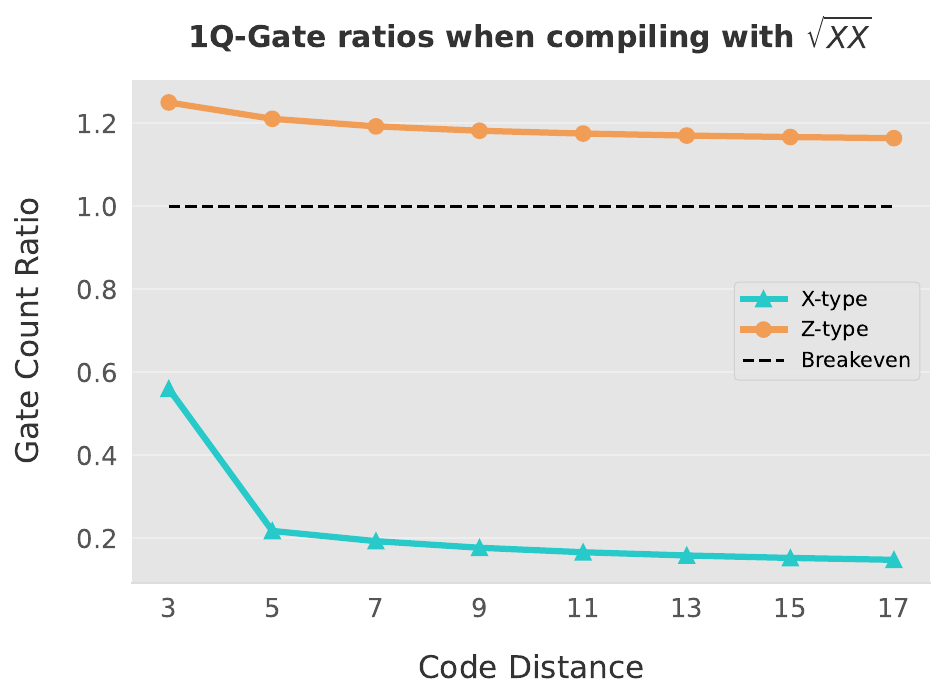}
	\end{subfigure}%
	\begin{subfigure}{.5\textwidth}
		\centering
		\includegraphics[scale=.5]{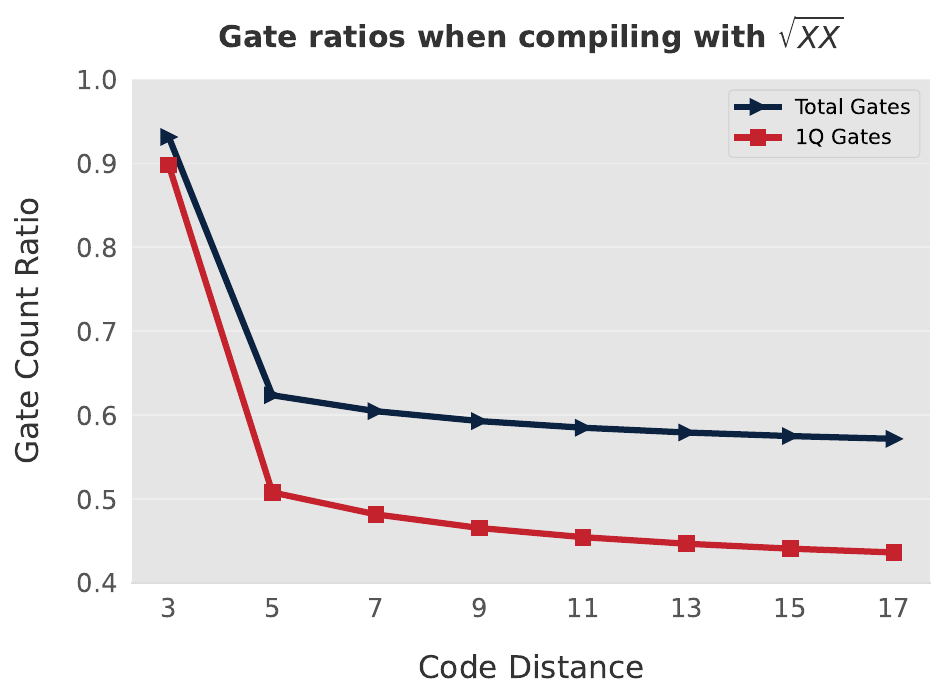}
	\end{subfigure}
	\caption{\emph{Left}: Ratio of single qubit gates present in the circuit when compiling with Tableaux Manipulation against Qiskit. Ratios are split into $X$-type (orange circles) and $Z$-type (blue triangles) single qubit gates. \emph{Right}: Ratio of the total gates (navy arrowheads) and total single qubit gates (red squares) when compiling with Tableaux Manipulation against Qiskit.}
	\label{fig:sqrtxx_compare}
\end{figure*}
\subsection{Comparison to open-source software}\label{sec:open_source}
Here we provide some comparisons to the Qiskit compiler functionality \cite{Qiskit}. We will be comparing a single round of syndrome extraction on a distance $d$ rotated surface code. For all code distances this circuit has depth-6 where four layers of \CX gates are sandwiched by a layer of Hadamards on either side. The circuits are generated with \texttt{stim} \cite{gidney2021stim} using the \texttt{stim.Circuit.generated} functionality, with kwargs \texttt{code\_task=surface\_code:rotated\_memory\_z, distance=$d$, rounds=1}. The circuit is identical regardless if the memory basis is chosen as $X$ or $Z$.

We compare compilation procedures using both the \sqrtXX gate and the $ECR$ gate. When compiling with the Tableaux Manipulation method, compilation was performed on \texttt{stim.Circuit} objects using $S$ and $\sqrt{X}$ as single qubit gates. We do not include frame-fixing Paulis when compiling with TM. When compiling with Qiskit, we convert the circuit from \texttt{stim.Circuit} into \texttt{qiskit.QuantumCircuit}, and call the \texttt{qiskit.compiler.transpile} function. The basis gates provided are \texttt{basis\_gates=[`rxx', `rx', `rz']} for \sqrtXX compilation and \texttt{basis\_gates=[`ecr', `rx', `rz']} for $ECR$ compilation.

We permit arbitrary angle rotations in Qiskit as we found this provided the most compact compilation. Manual checks were performed to ensure all angles were within tolerance to $\pm \pi$ or $\pm\pi/2$, making the gate Clifford. Where possible the \texttt{optimization\_level} kwarg was set to 3, however when compiling with \sqrtXX on distances $d=5$ and above this had to be dropped to 
\texttt{optimization\_level=1} as non-Clifford gates were being introduced. By explicitly setting the \texttt{basis\_gates} kwarg to be a list of Clifford operations, i.e. \texttt{[`z', `s', `sdg', `x', `sx', `sxdg', `ecr' / `rxx']}, the compiled circuit had significantly more gates (in the case of \texttt{ecr}) or raised an error (\texttt{rxx}).

\subsubsection{\sqrtXX}
To provide insight on the gate distributions of differently compiled circuits, we start by displaying a bar chart of the gates present for a $d=11$ rotated surface code syndrome extraction circuit, Fig.~\ref{fig:histogram_sqrtxx}. The uncompiled circuit consists of 440 \CX gates and 120 Hadamard gates. As Qiskit has access to arbitrary rotations in both $X$ and $Z$ bases it introduces a mix of $S$, $S^\dagger$ and $Z$ (and their equivalent Pauli-$X$ rotations). Tableaux Manipulation only introduces either $S$ or $\sqrt{X}$. For both compiled circuits the number of two-qubit gates is 440 as expected. TM also introduces marginally more $S$ gates than Qiskit, but significantly fewer $\sqrt{X}$ gates. 

To quantify this difference as the circuit size changes, in Fig.~\ref{fig:sqrtxx_compare} we plot the ratio of gate counts when compiling with TM against Qiskit. In the left hand plot we are displaying a breakdown of the single qubit Cliffords present in both circuits. Splitting them into their respective $X$-type and $Z$-type rotations, we can see that TM is consistently introducing more $S$ gates than Qiskit for all code distances: $125\%$ of the Qiskit count for $d=3$ and decreasing towards 116\% as the distance increases. However, TM introduces far fewer $\sqrt{X}$ gates than Qiskit: 56\% of the Qiskit count for $d=3$ and trending to below $15\%$ for increasing code distance. For some hardware architectures, such as superconducting qubits, $Z$-type rotations can be implemented `virtually'  making them effectively noiseless \cite{VirtualZGate}. The right hand plot compares both the single qubit gate count and the total gate count ratios. The first point in both curves representing $d=3$ is an outlier, as the circuit is quite small. For higher distance values, the single qubit gate count in a TM compiled circuit approaches 43\% of that in one compiled by Qiskit. In terms of total number of quantum operations, TM only uses 57\% of those used by Qiskit for a $d=17$ syndrome extraction circuit.

With regards to circuit depth, for both Qiskit and TM-compiled circuits across all code distances, the resulting circuits had 14 layers of gates.

\subsubsection{ECR}
\begin{figure}[t]
	\centering
	\includegraphics[width=\columnwidth]{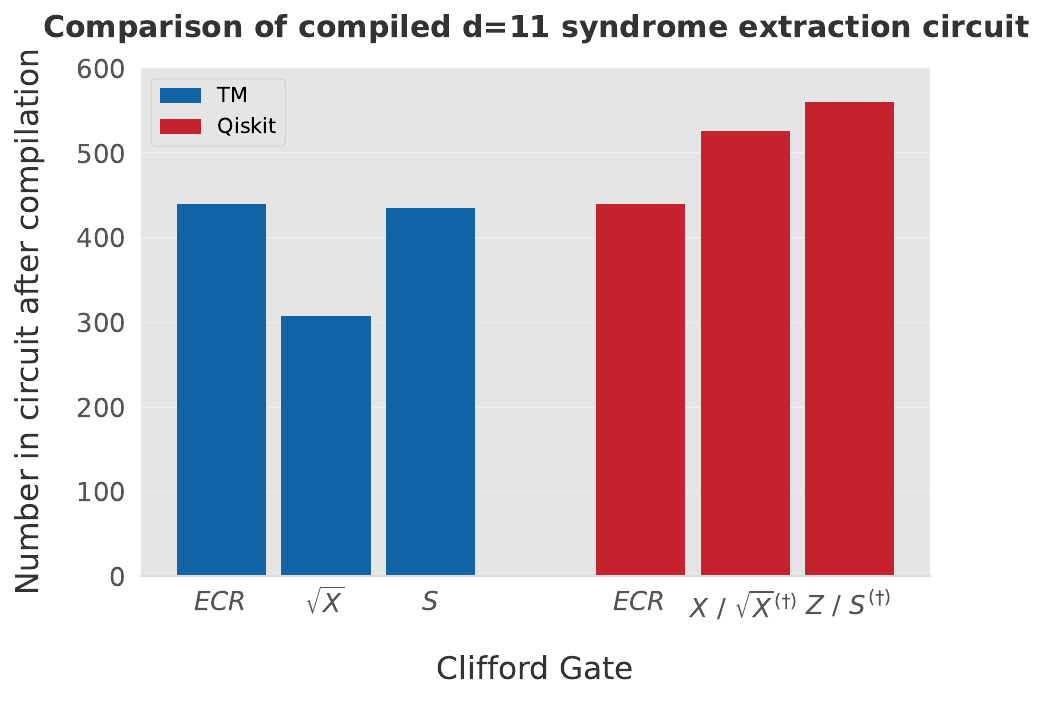}
 	\caption{Gate distribution when compiling a syndrome extraction circuit on a rotated surface code of distance $d=11$. The original circuit is 6 layers of gates, defined in terms of $H$ and \CX. Tableaux Manipulation (left, dark blue) and Qiskit (right, dark red) both compiled into a gateset with $ECR$ as the native two-qubit gate.}
	\label{fig:histogram_ecr}
\end{figure}
The $ECR$ gate, while Clifford, is not a native gate to \texttt{stim}. When compiling with this gate in \texttt{stim} we used the equivalent circuit in alternative Clifford gates:
\begin{eqnarray}\label{circuit:ecr}	
	\vcenter{
		\Qcircuit @C=.5em @R=.5em @!R {
			\lstick{} & \gate{{ECR}_{0}}\qwx[1] & \qw \\ 
			\lstick{} & \gate{ECR_1} & \qw }}
	\sp = \sp
	\vcenter{
	\Qcircuit @C=.5em @R=.5em @!R {
		\lstick{} & \gate{S} & \ctrl{1} & \gate{X}\\ 
		\lstick{} & \gate{\sqrt{X}} & \targ & \qw }}
\end{eqnarray}Where the subscripts $0,1$ in the $ECR$ gate diagram represents the role of the qubits, as the gate is antisymmetric. The tableau for $ECR$ is shown in  Eq.~(\ref{tableau:ecr}). This circuit was treated as a single two-qubit gate and gates were only added before or after $ECR$ layers. 

\begin{figure*}[t]
	\centering
	\begin{subfigure}{.5\textwidth}
		\centering
		\includegraphics[scale=.5]{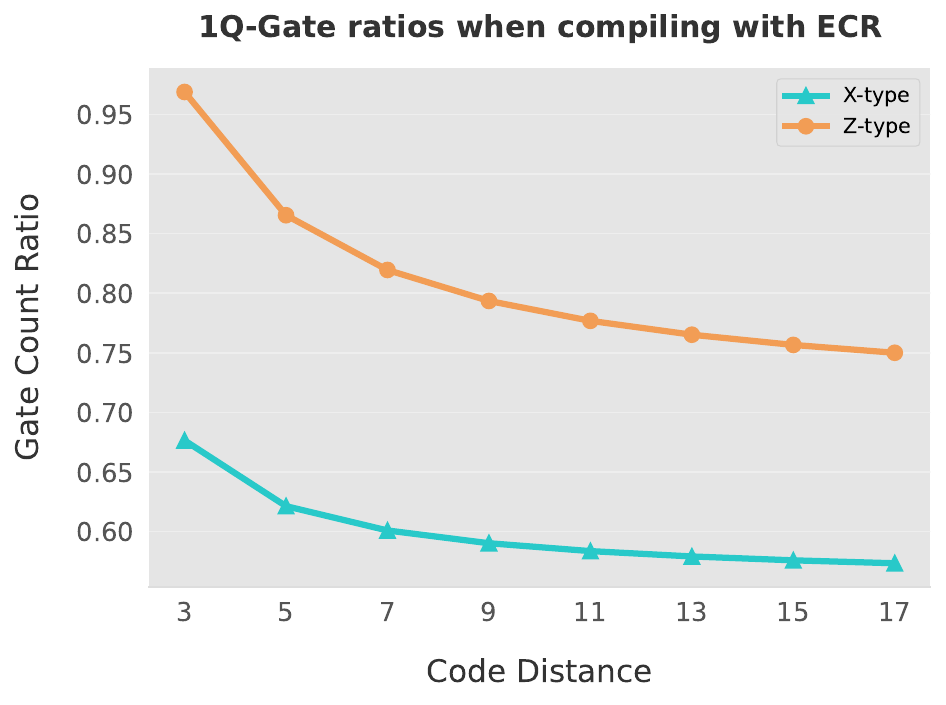}
		\label{fig:1q_gate_ratios_ecr}
	\end{subfigure}%
	\begin{subfigure}{.5\textwidth}
		\centering
		\includegraphics[scale=.5]{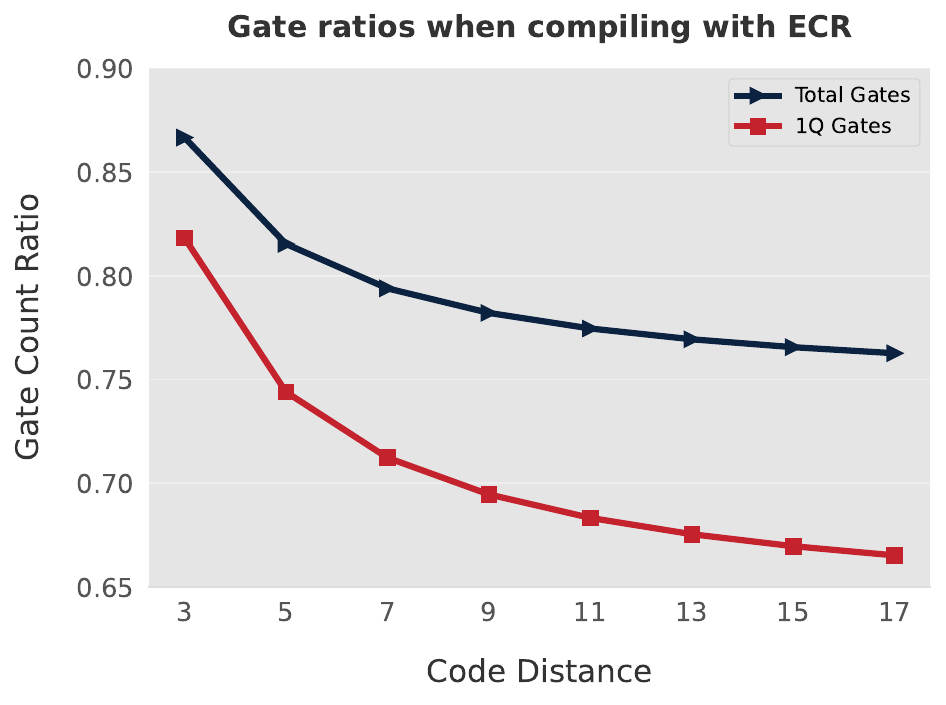}
		\label{fig:gate_ratios_ecr}
	\end{subfigure}
	\caption{\emph{Left}: Ratio of single qubit gates present in the circuit when compiling with Tableaux Manipulation against Qiskit. Ratios are split into $X$-type (orange circles) and $Z$-type (blue triangles) single qubit gates. \emph{Right}: Ratio of the total gates (navy arrowheads) and total single qubit gates (red squares) when compiling with Tableaux Manipulation against Qiskit.}
	\label{fig:ecr_compare}
\end{figure*}
In  Fig.~\ref{fig:histogram_ecr} we display the equivalent bar chart as in Fig.~\ref{fig:histogram_sqrtxx} for compilation with an $ECR$ gate. While the effect is perhaps more muted than that of \sqrtXX, TM is consistently introducing fewer single qubit gates than Qiskit. This reduction is quantified across all code distances in Fig.~\ref{fig:ecr_compare}, where we plot the ratio of gate counts of TM against Qiskit. Across all code distances we introduce fewer single qubit gates than Qiskit, with the effect more pronounced for $X$-type rather than $Z$-type gates. For $Z$-type gates, we reduce the total count down to $75\%$ of Qiskits total for $d=17$; for $X$-type gates TM introduces only 57\% of gates compared to Qiskit.

Looking at the right hand plot, the ratio of single qubit gates introduced by TM against Qiskit drops from $<82\%$ for the smallest code size and approaches $66\%$ for the highest distance considered.
When compiling with an $ECR$ gate for $d=17$, a circuit compiled by TM will only have about $76\%$ of the gates compared to one compiled by Qiskit. 

With regards to circuit depth we take into account that when compiling with the circuit equivalency shown in Eq.~(\ref{circuit:ecr}) a single layer is being represented by three layers. Correcting for this, across all code distances for both Qiskit and TM-compiled circuits, the compiled circuits had 15 layers of gates.

Qiskit was the best performing open-source compiler we tested. A future version of this paper may include comparisons to other compilers. We compare favorably to Qiskit as it is \emph{not} a Clifford compiler: it is designed for arbitrary quantum gates. As mentioned, restricting Qiskit to only use Clifford operations resulted in either a failure or degrading performance, indicating that this process (or one similar to it) is not being employed. 


\subsection{Implications for logical error}
Here we will briefly highlight how the above results affect the logical error probability of quantum memory experiments. We present the results for a $d=11$ quantum memory being compiled into a gateset with \sqrtXX. As this gate is native to \texttt{stim} it is straightforward for noisy simulations. For context, at this distance TM uses about 58.5\% of the quantum operations compared to Qiskit (see Fig.~\ref{fig:sqrtxx_compare}).

The circuits are generated with \texttt{stim} as described in Sec.~\ref{sec:open_source}, setting \texttt{distance=11} and \texttt{rounds=11}. The syndrome extraction subcircuit (4 layers of \CX sandwiched by 2 layers of $H$) is extracted and individually compiled using both TM and Qiskit into a gateset with \sqrtXX as the two-qubit gate. The single qubit gates used for compilation are as in Sec.~\ref{sec:open_source}. The compiled circuits are then reinserted into the \texttt{stim} circuit in place of the $H+CX$ subcircuit, such that we have three copies of the same $d=11$ quantum memory experiment defined in \texttt{stim}: one using $H+CX$, one compiled by TM and one compiled by Qiskit. 

We apply a depolarizing noise model to each circuit, such that:
\begin{itemize}
    \item Two-qubit gates have a two-qubit depolarizing channel applied with probability $p$.
    \item Single qubit gates have a single qubit depolarizing channel applied with probability $p/10$.
    \item Reset operations in the $Z$ basis have a single qubit depolarizing channel applied with probability $p/10$.
    \item Measurement outcomes are flipped with probability $p$.
    \item For each layer in the circuit, any idle qubits have a single qubit depolarizing noise channel applied with probability $p/10$.  
\end{itemize}
Decoding was then performed using the \texttt{PyMatching} package \cite{higgott2021pymatching}, with each circuit being sampled with $100p^{-2}$ shots. We ran simulations for varying noise values around the threshold, which for this noise model on logical-$Z$ memory using a $H+CX$ gateset is $0.00927(9)$. When compiling with Qiskit, the threshold drops to  $0.00747(23)$; for TM the threshold is $0.00797(54)$. The results for the $d=11$ case are reported in Table~\ref{tab:logical_error_values}. 

We see that compiling into a native gateset causes the logical error probability to jump significantly, and that this effect is suppressed by using Tableaux Manipulation. At $p=0.01$, we see that TM is reducing the logical error probability by about $\sim11\%$, and this reduction improves as the physical error decreases below threshold. At $p=0.004$, we are achieving $\sim18\%$ fewer logical errors by compiling with TM versus Qiskit. Beyond this point the TM / Qiskit ratio begins to rise again, as we enter the regime where single qubit gates become essentially noiseless in comparison to two-qubit gates and measurements. Far below threshold, say at $p=10^{-4}$, it is likely that the logical error probability would be equivalent across all three circuit implementations. However, we argue that it is still advantageous to employ TM as communicating instructions to physical qubits still entails a non-zero heat cost, and TM allows you to achieve the same circuit implementations with fewer quantum operations. 
\begin{table*}[]
    \centering
    \renewcommand{\arraystretch}{1.25}
    \begin{tabular}{c|ccc|c}
         Depolarizing\\Noise $p$ & $H$+\CX & TM Compiled & Qiskit Compiled & TM / Qiskit  \\
         \hline
         $0.002$ & $(6.12\pm 2.81)\times 10^{-6}$ &  $2.07(43)\times 10^{-5}$ &  $2.49(41)\times 10^{-5}$ &$0.846(187)$\\
         $0.004$ & $4.13(42)\times 10^{-4}$ &  $1.39(8)\times 10^{-3}$&  $1.70(7)\times 10^{-3}$ & $0.819(48)$\\
         $0.006$ & $4.68(24)\times 10^{-3}$ &  $1.41(3)\times 10^{-2   }$&  $1.7(0)\times 10^{-2}$ & $0.828(25)$\\
         $0.008$ & $0.023(1)$  & $0.06(1)$& $0.07(1)$&$0.851(15)$\\
         $0.01$ & 0.0683(12) & 0.152(2) & 0.172(2) & $0.886(13)$
    \end{tabular}
    \caption{Logical error values for a $d=11$ quantum memory experiment experiencing a depolarizing noise model decoded with minimum weight perfect matching. We compare the original circuit defined in terms of $H+CX$ to a circuit compiled into a gateset using \sqrtXX with both TM and Qiskit. }
    \label{tab:logical_error_values}
\end{table*}
\section{Discussion}\label{sec:discussion}
We have presented a simple, intuitive method for compiling a Clifford interaction into another by running a find-and-replace protocol for entangling operations, and fixing up the instantaneous Pauli conjugation as necessary. In comparing to open-source alternatives, compilation of syndrome extraction circuits for the rotated surface code resulted in a reduction of required quantum operations across all code distances. 

For the class of circuits we have considered, experimentally this method provides a net positive with no clear downsides. For near-term devices exhibiting a noise profile roughly around $1\%$ depolarizing, this method of compilation will improve logical fidelity in experiments. When hardware improves to the point that single qubit gates are effectively noiseless, this protocol maintains an advantage over existing compilers by requiring fewer quantum operations communicated to physical qubits.
 
Practicality aside, the procedure outlined above is useful for building intuition around Clifford gates and circuits, and accelerating certain aspects of research when dealing with alternative Clifford gatesets beyond \clifford. Directly dealing with tableaux allows extraneous detail of the Clifford interaction to be abstracted away and reduced to ensuring the conjugation is correct.

We have only considered circuits employing one kind of entangling Clifford, and focused on compiling into gatesets using the same kind of entangling Clifford; we provide some discussion about replacing a \CX-like gate with an \iSWAP-like gate in Appendix \ref{iswap_discussion}. However, this is insufficient to determine how this process would perform on arbitrary Clifford circuits. Given a gateset with both \CX- and \iSWAP-like Cliffords, the initial condition can be easily adapted to replacing each entangling gate in an arbitrary circuit with its corresponding type in our gateset. If we only have access to one type of entangling Clifford, compiling an arbitrary stabilizer circuit would likely require further algorithms to ensure the minimal number of entangling gates is employed, such as that in Ref.~\cite{Bravyi2021}. 

\bibliographystyle{unsrtnat}
\bibliography{tm_bib}

\begin{thebibliography}{19}
\providecommand{\natexlab}[1]{#1}
\providecommand{\url}[1]{\texttt{#1}}
\expandafter\ifx\csname urlstyle\endcsname\relax
  \providecommand{\doi}[1]{doi: #1}\else
  \providecommand{\doi}{doi: \begingroup \urlstyle{rm}\Url}\fi

\bibitem[Litinski(2019)]{Litinski2019gameofsurfacecodes}
Daniel Litinski.
\newblock A {G}ame of {S}urface {C}odes: {L}arge-{S}cale {Q}uantum {C}omputing with {L}attice {S}urgery.
\newblock \emph{{Quantum}}, 3:\penalty0 128, March 2019.
\newblock ISSN 2521-327X.
\newblock \doi{10.22331/q-2019-03-05-128}.
\newblock URL \url{https://doi.org/10.22331/q-2019-03-05-128}.

\bibitem[Horsman et~al.(2012)Horsman, Fowler, Devitt, and Meter]{Horsman_2012}
Dominic Horsman, Austin~G Fowler, Simon Devitt, and Rodney~Van Meter.
\newblock Surface code quantum computing by lattice surgery.
\newblock \emph{New Journal of Physics}, 14\penalty0 (12):\penalty0 123011, dec 2012.
\newblock \doi{10.1088/1367-2630/14/12/123011}.
\newblock URL \url{https://dx.doi.org/10.1088/1367-2630/14/12/123011}.

\bibitem[AI(2023)]{GoogleScaling}
Google~Quantum AI.
\newblock Suppressing quantum errors by scaling a surface code logical qubit.
\newblock \emph{Nature}, 614\penalty0 (7949):\penalty0 676--681, Feb 2023.
\newblock ISSN 1476-4687.
\newblock \doi{10.1038/s41586-022-05434-1}.
\newblock URL \url{https://doi.org/10.1038/s41586-022-05434-1}.

\bibitem[Moses~et al.(2023)]{Quantinuum}
S.~A. Moses~et al.
\newblock A race-track trapped-ion quantum processor.
\newblock \emph{Phys. Rev. X}, 13:\penalty0 041052, Dec 2023.
\newblock \doi{10.1103/PhysRevX.13.041052}.
\newblock URL \url{https://link.aps.org/doi/10.1103/PhysRevX.13.041052}.

\bibitem[Bluvstein~et al.(2024)]{neutralatoms}
Dolev Bluvstein~et al.
\newblock Logical quantum processor based on reconfigurable atom arrays.
\newblock \emph{Nature}, 626\penalty0 (7997):\penalty0 58--65, Feb 2024.
\newblock ISSN 1476-4687.
\newblock \doi{10.1038/s41586-023-06927-3}.
\newblock URL \url{https://doi.org/10.1038/s41586-023-06927-3}.

\bibitem[M\o{}lmer and S\o{}rensen(1999)]{MSgate}
Klaus M\o{}lmer and Anders S\o{}rensen.
\newblock Multiparticle entanglement of hot trapped ions.
\newblock \emph{Phys. Rev. Lett.}, 82:\penalty0 1835--1838, Mar 1999.
\newblock \doi{10.1103/PhysRevLett.82.1835}.
\newblock URL \url{https://link.aps.org/doi/10.1103/PhysRevLett.82.1835}.

\bibitem[Malekakhlagh et~al.(2020)Malekakhlagh, Magesan, and McKay]{ECRGate}
Moein Malekakhlagh, Easwar Magesan, and David~C. McKay.
\newblock First-principles analysis of cross-resonance gate operation.
\newblock \emph{Phys. Rev. A}, 102:\penalty0 042605, Oct 2020.
\newblock \doi{10.1103/PhysRevA.102.042605}.
\newblock URL \url{https://link.aps.org/doi/10.1103/PhysRevA.102.042605}.

\bibitem[Arute~et al.(2019)]{Arute2019}
Frank Arute~et al.
\newblock Quantum supremacy using a programmable superconducting processor.
\newblock \emph{Nature}, 574\penalty0 (7779):\penalty0 505--510, Oct 2019.
\newblock ISSN 1476-4687.
\newblock \doi{10.1038/s41586-019-1666-5}.
\newblock URL \url{https://doi.org/10.1038/s41586-019-1666-5}.

\bibitem[Bravyi and Maslov(2021)]{Bravyi_2021}
Sergey Bravyi and Dmitri Maslov.
\newblock Hadamard-free circuits expose the structure of the clifford group.
\newblock \emph{IEEE Transactions on Information Theory}, 67\penalty0 (7):\penalty0 4546–4563, July 2021.
\newblock ISSN 1557-9654.
\newblock \doi{10.1109/tit.2021.3081415}.
\newblock URL \url{http://dx.doi.org/10.1109/TIT.2021.3081415}.

\bibitem[van~den Berg(2021)]{berg2021simple}
Ewout van~den Berg.
\newblock A simple method for sampling random clifford operators, 2021.
\newblock quant-ph/2008.06011.

\bibitem[Bravyi et~al.(2021)Bravyi, Shaydulin, Hu, and Maslov]{Bravyi2021}
Sergey Bravyi, Ruslan Shaydulin, Shaohan Hu, and Dmitri Maslov.
\newblock Clifford {C}ircuit {O}ptimization with {T}emplates and {S}ymbolic {P}auli {G}ates.
\newblock \emph{{Quantum}}, 5:\penalty0 580, November 2021.
\newblock ISSN 2521-327X.
\newblock \doi{10.22331/q-2021-11-16-580}.
\newblock URL \url{https://doi.org/10.22331/q-2021-11-16-580}.

\bibitem[Kliuchnikov and Maslov(2013)]{Maslov2013}
Vadym Kliuchnikov and Dmitri Maslov.
\newblock Optimization of clifford circuits.
\newblock \emph{Phys. Rev. A}, 88:\penalty0 052307, Nov 2013.
\newblock \doi{10.1103/PhysRevA.88.052307}.
\newblock URL \url{https://link.aps.org/doi/10.1103/PhysRevA.88.052307}.

\bibitem[Gottesman and Chuang(1999)]{Gottesman1999}
Daniel Gottesman and Isaac~L. Chuang.
\newblock Demonstrating the viability of universal quantum computation using teleportation and single-qubit operations.
\newblock \emph{Nature}, 402\penalty0 (6760):\penalty0 390--393, Nov 1999.
\newblock ISSN 1476-4687.
\newblock \doi{10.1038/46503}.
\newblock URL \url{https://doi.org/10.1038/46503}.

\bibitem[Gottesman(1997)]{gottesman1997stabilizer}
Daniel Gottesman.
\newblock Stabilizer codes and quantum error correction.
\newblock 1997.
\newblock quant-ph/9705052.

\bibitem[Aaronson and Gottesman(2004)]{Aaronson_2004}
Scott Aaronson and Daniel Gottesman.
\newblock Improved simulation of stabilizer circuits.
\newblock \emph{Physical Review A}, 70\penalty0 (5), November 2004.
\newblock ISSN 1094-1622.
\newblock \doi{10.1103/physreva.70.052328}.
\newblock URL \url{http://dx.doi.org/10.1103/PhysRevA.70.052328}.

\bibitem[Gidney(2021)]{gidney2021stim}
Craig Gidney.
\newblock Stim: a fast stabilizer circuit simulator.
\newblock \emph{{Quantum}}, 5:\penalty0 497, July 2021.
\newblock ISSN 2521-327X.
\newblock \doi{10.22331/q-2021-07-06-497}.
\newblock URL \url{https://doi.org/10.22331/q-2021-07-06-497}.

\bibitem[{Qiskit contributors}(2023)]{Qiskit}
{Qiskit contributors}.
\newblock Qiskit: An open-source framework for quantum computing, 2023.
\newblock 10.5281/zenodo.2573505.

\bibitem[McKay et~al.(2017)McKay, Wood, Sheldon, Chow, and Gambetta]{VirtualZGate}
David~C. McKay, Christopher~J. Wood, Sarah Sheldon, Jerry~M. Chow, and Jay~M. Gambetta.
\newblock Efficient z-gates for quantum computing.
\newblock \emph{Physical Review A}, 96\penalty0 (2), August 2017.
\newblock ISSN 2469-9934.
\newblock \doi{10.1103/physreva.96.022330}.
\newblock URL \url{http://dx.doi.org/10.1103/PhysRevA.96.022330}.

\bibitem[Higgott(2021)]{higgott2021pymatching}
Oscar Higgott.
\newblock Pymatching: A python package for decoding quantum codes with minimum-weight perfect matching, 2021.
\newblock quant-ph/2105.13082.

\end{thebibliography}

\appendix
\section{Caveats for the broader Clifford class}\label{iswap_discussion}
\begin{align}\label{tableau:xxxx}
	&\nonumber\vcenter{
		\Qcircuit @C=.5em @R=.5em @!R {
		\lstick{0} & \gate{H} & \ctrl{1} & \ctrl{2} & \ctrl{3} & \ctrl{4} & \gate{H} & \qw \\
		\lstick{1} & \qw & \targ & \qw & \qw &\qw &\qw &\qw \\
		\lstick{2} & \qw & \qw & \targ & \qw &\qw &\qw &\qw\\
		\lstick{3} & \qw & \qw & \qw & \targ &\qw &\qw &\qw\\
		\lstick{4} & \qw & \qw & \qw & \qw &\targ &\qw &\qw\\
		}
	}
	\\
	\\
	&\sp=\nonumber\begin{tabular}{c|ccccc}
		& $X_0~Z_0$ & $X_1~Z_1$ & $X_2~Z_2$ & $X_3~Z_3$ & $X_4~Z_4$\\
		\hline
		$\pm$ & $+~+$ & $+~+$ & $+~+$& $+~+$& $+~+$\\
		0 & $X~Z$ & $ \undr~X$ & $ \undr~X$ & $ \undr~X$ & $ \undr~X$\\
		1 & $\undr~X$ & $X~Z$ & $\undr~\undr$ & $\undr~\undr$  & $\undr~\undr$ \\
		2 & $\undr~X$ & $\undr~\undr$ & $X~Z$ & $\undr~\undr$  & $\undr~\undr$ \\
		3 & $\undr~X$ & $\undr~\undr$  & $\undr~\undr$ & $X~Z$ & $\undr~\undr$ \\
		4 & $\undr~X$ & $\undr~\undr$  & $\undr~\undr$  & $\undr~\undr$& $X~Z$ \\
	\end{tabular}.
\end{align}
In the main text we are focused on circuits defined on \CX-like entangling Cliffords being compiled into a gateset which also uses \CX-like entangling Cliffords. The same process can be directly applied to circuits defined on only \iSWAP-like entangling Cliffords, being compiled into a gateset which also uses \iSWAP-like entangling Cliffords. As the protocol relies on a heuristic of replacing entangling gates with entangling gates (implicitly between the same qubits), it applies equally to both these cases. However, if we wish to compile a circuit defined on a \CX-like gate into a gateset that only uses an \iSWAP-like gate, or vice versa, this heuristic breaks down.

Consider the weight-4 $X$-stabiliser circuit shown in Eq.~(\ref{tableau:xxxx}). If we replace each \CX gate with an \iSWAP between the same qubits (such that qubit 0 is involved in four \iSWAP operations) we will not be within single qubit rotations to the desired Clifford interaction. Due to the swapping aspect of these gates we must be aware of which qubit states have been swapped, tracking an `instantaneous qubit index' so to speak. Consider the following as an alternative initial circuit for TM compilation: 
\begin{eqnarray}\label{tableau:iswap_xxxx}
	&\nonumber\vcenter{
		\Qcircuit @C=.5em @R=.5em @!R {
			\lstick{0} &  \gate{iSW\!AP}\qwx[1] &\qw & \qw & \qw &\qw\\ 
			\lstick{1} & \gate{iSW\!AP} & \gate{iSW\!AP}\qwx[1]& \qw & \qw & \qw\\ 
			\lstick{2} & \qw & \gate{iSW\!AP} & \gate{iSW\!AP}\qwx[1] & \qw& \qw\\ 
			\lstick{3} & \qw & \qw & \gate{iSW\!AP} & \gate{iSW\!AP}\qwx[1]& \qw\\ 
			\lstick{4} & \qw & \qw &\qw & \gate{iSW\!AP}& \qw\\ 
		}
	}
	 \\ 
	 \\ 
	&\sp=\sp\nonumber\begin{tabular}{c|ccccc}
		& $X_0~Z_0$ & $X_1~Z_1$ & $X_2~Z_2$ & $X_3~Z_3$ & $X_4~Z_4$\\
		\hline
		$\pm$ & $+~+$ & $+~+$ & $+~+$& $+~+$& $+~+$\\
		0 & $Z~\undr$ & $ Y~Z$ & $ \undr~\undr$ & $ \undr~\undr$ & $ \undr~\undr$\\
		1 & $Z~\undr$ & $\undr~\undr$ & $Y~Z$ & $\undr~\undr$  & $\undr~\undr$ \\
		2 & $Z~\undr$ & $\undr~\undr$ & $\undr~\undr$ & $Y~Z$  & $\undr~\undr$ \\
		3 & $Z~\undr$ & $\undr~\undr$  & $\undr~\undr$ & \undr~\undr & $Y~Z$ \\
		4 & $X~Z$ & $Z~\undr$  & $Z~\undr$  & $Z~\undr$& $Z~\undr$ \\
	\end{tabular}
\end{eqnarray}
This circuit can be made equivalent to Eq.~(\ref{tableau:xxxx}) using only single qubit Clifford gates, up to some permutation of rows. Qubit 0 is no longer the qubit to be measured, and has been swapped into qubit 4. This is a relaxation of the requirement that two stabilizer tableaux be equal to ensure compilation, but comes at the cost of qubit state shuffling. 
Indeed, on certain qubit architectures the connectivity graph dictates that the circuit in Eq.~(\ref{tableau:xxxx}) is valid and the circuit in Eq.~(\ref{tableau:iswap_xxxx}) is not. In this case we can then either a) introduce further qubits to the tableaux to act as intermediaries, wherein our goal is to replicate the target tableau in a \emph{subset} of rows in the new tableau, up to some permutation, or b) introduce further \iSWAP and single qubit gates within the original set of qubits to undo the state swapping.

In the cases where an informed initial condition is not immediately obvious, we propose employing the initial condition as normal and then inserting $SW\!AP$ gates after each \iSWAP, between the same qubits, such that $\left(iSW\!AP+SW\!AP\right)$ approximates a \CX. Compilation can then proceed as normal. Once the tableaux are equivalent, iterate through the circuit removing each $SW\!AP(i,~j)$ gate sequentially, and updating following qubit indices ($i\leftrightarrow j$) as necessary. This will not correct for all possible edge cases, such as the restriction in connectivity mentioned above.

\end{document}